\documentclass{aastex62}
\graphicspath{{./}{figures/}}
\shortauthors{Wang et al.}

\begin{document}
\title{Magnetic Structure of an Erupting Filament}

\author[0000-0001-5589-0416]{Shuo Wang}
\affil{Department of Astronomy, New Mexico State University, P.O. Box 30001, MSC 4500, Las Cruces, NM 88003, USA}
\affil{DKIST Ambassador}

\author[0000-0002-8975-812X]{Jack M. Jenkins}
\affiliation{Mullard Space Science Laboratory, University College London, Holmbury St. Mary, Dorking, Surrey, RH5 6NT, UK}

\author{Valentin Martinez Pillet}
\affiliation{National Solar Observatory, 3665 Discovery Drive, Boulder, CO 80303, USA}

\author[0000-0001-7706-4158]{Christian Beck}
\affiliation{National Solar Observatory, 3665 Discovery Drive, Boulder, CO 80303, USA}

\author[0000-0003-3137-0277]{David M. Long}
\affiliation{Mullard Space Science Laboratory, University College London, Holmbury St. Mary, Dorking, Surrey, RH5 6NT, UK}

\author[0000-0002-9308-3639]{Debi Prasad Choudhary}
\affiliation{Department of Physics and Astronomy, California State University Northridge, Northridge, CA 91330-8268, USA}

\author[0000-0002-5547-9683]{Karin Muglach}
\affiliation{Catholic University of America, Washington, DC 20064, USA}
\affiliation{NASA Goddard Space Flight Center, Greenbelt, MD 20771, USA}

\author[0000-0003-1493-101X]{James McAteer}
\affil{Department of Astronomy, New Mexico State University, P.O. Box 30001, MSC 4500, Las Cruces, NM 88003, USA}
\affil{Sunspot Solar Observatory, Sunspot, NM 88349, USA}

\correspondingauthor{James McAteer}
\email{mcateer@nmsu.edu}

\begin{abstract}

The full 3-D vector magnetic field of a solar filament prior to eruption is presented. The filament was observed with the Facility Infrared Spectropolarimeter at the Dunn Solar Telescope in the chromospheric He~{\sc i} line at 10830~\AA\ on May 29 and 30, 2017. We inverted the spectropolarimetric observations with the HAnle and ZEeman Light (HAZEL) code to obtain the chromospheric magnetic field. A bimodal distribution of field strength was found in or near the filament. The average field strength was 24 Gauss, but prior to the eruption we find the 90th percentile of field strength was 435 Gauss for the observations on May 29. The field inclination was about 67 degree from the solar vertical. The field azimuth made an angle of about 47 to 65 degree to the spine axis. The results suggest an inverse configuration indicative of a flux rope topology. He~{\sc i} intensity threads were found to be co-aligned with the magnetic field direction. The filament had a sinistral configuration as expected for the southern hemisphere. The filament was stable on May 29, 2017 and started to rise during two observations on May 30, before erupting and causing a minor coronal mass ejection. There was no obvious change of the magnetic topology during the eruption process. Such information on the magnetic topology of erupting filaments could improve the prediction of the geoeffectiveness of solar storms.

\end{abstract}

\keywords{Sun: filaments, prominences --- Sun: infrared--- Sun: magnetic fields}

\section{Introduction} \label{sec:intro}
Solar filaments are relatively thin and elongated, magnetically bound structures observed in absorption when projected against the chromosphere of the solar disk. They are composed of cold ($10^4$ K), dense (electron density $> 10^9~ $cm$^{-3}$) plasma which is thermally insulated from its much hotter surroundings in the transition region and corona \citep{2014LRSP...11....1P}. Filaments are always located above magnetic field polarity inversion lines in the photosphere. Depending on the magnetic conditions within the photosphere below, filaments may be categorized as either active, intermediate, or quiescent. Quiescent filaments are typically more extended (100s of Mm), elevated (10’s of Mm), and long-lived (months) than their active region counterparts, in addition to exhibiting weaker host magnetic field (a few Gauss) \citep[\textit{e.g.},][]{2010SSRv..151..333M}. Filaments are composed of fine threads which form an angle ($20^{\circ}$ -- $35^{\circ}$) to their spine \citep{2007SoPh..246...65L,2014LRSP...11....1P}. Filaments and prominences can be observed in strong chromospheric spectral lines such as H$\alpha$ at 6563~\AA , or the absorption lines of He~{\sc i} at 5876~\AA (He~{\sc i} D3) and 10830~\AA . In particular, spectropolarimetry of 10830~\AA\ is a powerful tool used to study the magnetic structure in the upper atmosphere \citep[e.g.][]{2002SoPh..209..349C,2013ApJ...768..111S,2015SoPh..290.1607S,2016ApJ...833....5S}. \citet{1998ApJ...493..978L} demonstrated measurements of the full Stokes parameters of a filament at 10830~\AA , and proposed a tilted magnetic field loop across the filament spine to explain their observation.

\citet{2003ApJ...598L..67C} presented results of He~{\sc i} D3 inversions in a prominence, finding a range of magnetic field strengths of between 10 - 70 Gauss. \citet{2006ApJ...642..554M} found magnetic field strengths of about 30 Gauss in a polar crown prominence using observations taken in the He~{\sc i} line at 10830~\AA .  However, much larger field strengths have also been observed. For example, an active region filament with a magnetic field strength of up to 800 Gauss was observed on May 17, 2005, by \cite{2012ApJ...749..138X}. The same filament was activated by a solar flare on May 18, 2005, and showed multiple components in the profiles \citep{2011A&A...526A..42S}. \cite{2014A&A...566A..46O} studied the magnetic field of a prominence, and found that the magnetic field was stronger in its feet than in its main body. However, the study of the magnetic field of prominences is not without its complications. For example, \citet{2017A&A...597A..31M} studied the multidimensional effects on the inference of the magnetic field of prominences, and found that the inferred magnetic field is weaker and more horizontal than their input. Furthermore and quite crucially, \citet{2019A&A...625A.128D,2019A&A...625A.129D} studied an active region filament observed on June 17, 2014, and found that the Stokes V signals were not from the filament, but from the active region below it.

In addition to the amplitude, inversions also provide information about the orientation of the filament’s host magnetic field. \citet{2003ASPC..307..468C} studied the azimuth of two filaments close to disk center, and found that it is oriented approximately along the filament axis. \citet{2007ASPC..368..347M} studied the orientation of the magnetic field of a solar filament near disk center, and found that the inclination of the magnetic field is horizontal in the central part of the filament and changes along the filament axis, and its azimuth changes along the filament as well. \citet{2017ApJ...851..130H} statistically studied the orientation of the magnetic field of filaments and confirmed their hemispheric pattern of chirality. 

After filament formation, filaments can be stable within the solar atmosphere for a few rotations.  The stability of filaments is heavily dependent on the evolution of its host magnetic field, in the case of quiescent filaments it is believed their stability is additionally dependent on the filament plasma itself \citep[\textit{see e.g.},][]{1998Natur.396..440Z,2005ApJ...630L..97T,2006PhRvL..96y5002K,2008ApJ...676L..89B,2018SoPh..293....7J,2018ApJ...862...54F,2019ApJ...873...49J}. When this stability is removed, some filament eruptions lead to coronal mass ejections \citep{2006LRSP....3....2S,2002A&A...395..257P}, while others do not \citep{2003GeoRL..30.2107C}. The magnetic structure of an active region filament observed on May 17, 2005 was explained as a rising flux rope by \citet{2014A&A...561A..98S}. \cite{2009A&A...501.1113K,2012A&A...539A.131K,2012A&A...542A.112K} studied an active region filament observed in July, 2005, and suggested a flux rope topology. Together, the field strengths and angles can be used to distinguish between the multiple topologies that were suggested to be able to support the filament plasma, such as the sheared arcade model and the flux rope model \citep{2018LRSP...15....7G}.

Regular full vector magnetic fields of solar filaments are now obtained from synoptic observations at the DST. In this manuscript we present the inversion of He~{\sc i} 10830~\AA\ as observed within a filament during an observing campaign carried out in May 2017. In addition, we present the first high-resolution observations in He~{\sc i} 10830 \AA\ acquired during a filament eruption along with the corresponding HAnle and ZEeman Light \cite[HAZEL;][]{2011ascl.soft09004A} inversion results. The line-of-sight velocity information will be used in a future paper. Section 2 describes our data sets. The analysis with the HAZEL code is detailed in Section 3. Our results are presented in Section 4 and discussed in Section 5. Section 6 gives our conclusions.

\section{Observations} \label{sec:obs}
\begin{figure}
\plotone{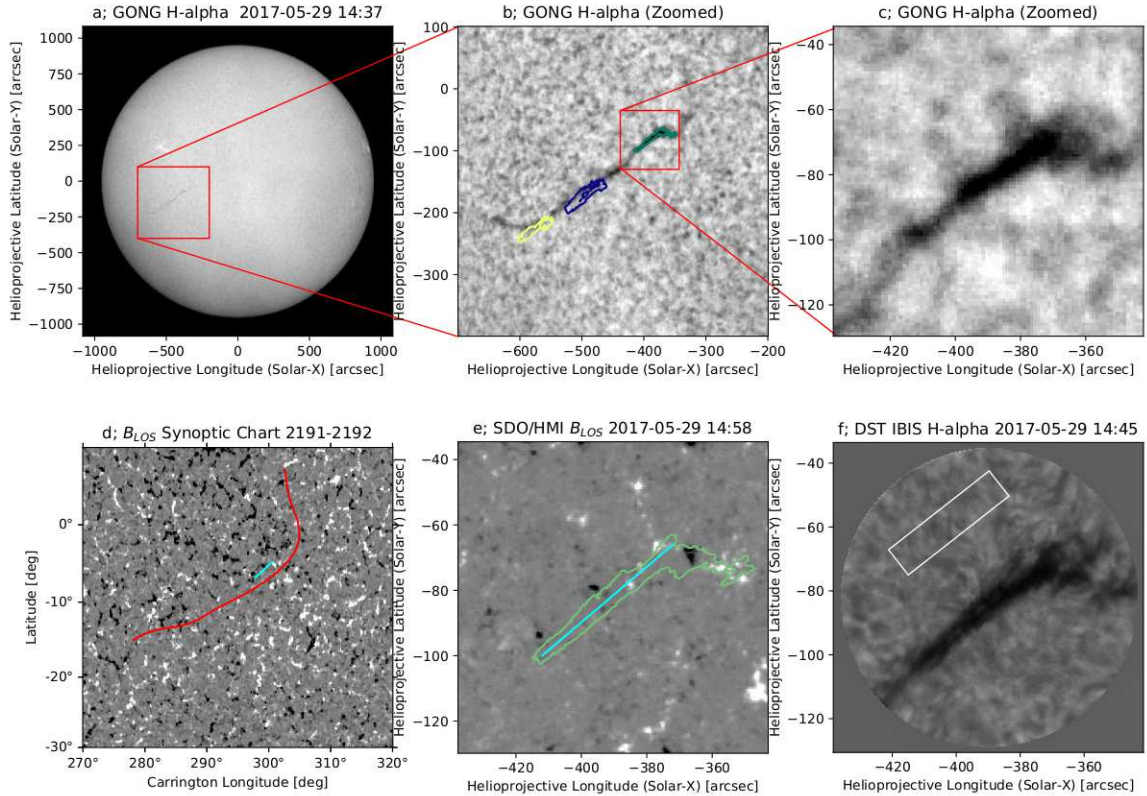}
\caption{Filament observed on May 29 2017. \textit{Panel a}; Context full-disk H$\alpha$ line-core image taken by the GONG instrument. The box indicates the location of the filament. \textit{Panel b}; Zoom-in of the box in panel a, showing the whole filament. Green, dark blue and yellow contours are the part of the filament observed by FIRS on May 29/30. \textit{Panel c}; Zoom-in of the box in panel b, showing the appearance of the hooked shape of the filament. \textit{Panel d}; Carrington synoptic map of SDO/HMI line-of-sight magnetic field. The red curve shows the polarity inversion line, and the cyan line at about (299, -6) degree shows the position of the filament axis observed by FIRS on May 29. The image is saturated at $\pm$20 Gauss. \textit{Panel e}; SDO/HMI  line-of-sight magnetogram. The cyan line in Panel d is shown in this panel as well. The image is saturated at $\pm$100 Gauss. \textit{Panel f}; The H$\alpha$ line-core image of the filament taken by IBIS. The white box indicates the quiet-Sun area over which the average IBIS profile was calculated. Images observed by FIRS are aligned to $\sim$1 arcsec.\label{fig:fig1}}
\end{figure}

A quiescent filament, shown in Figure 1, was observed on-disk (-24.3\degr,-4.85\degr) ([x: -500$''$, y: -200$''$]) at 14:44:55~UT on May 29 2017. Figure~\ref{fig:fig1}a shows a full-disk context image taken in the H$\alpha$ line core at 6562.8~\AA\ using the GONG network\footnote{\href{http://halpha.nso.edu/index.html}{http://halpha.nso.edu/index.html}} \citep{2011SPD....42.1745H}. Figure~\ref{fig:fig1}b is a zoom-in of the box in Figure~\ref{fig:fig1}a showing the filament of interest, this is then further zoomed to the FOV of the ground-based instruments at the DST in Figure~\ref{fig:fig1}c. Simultaneous photospheric magnetic field information (Figure~\ref{fig:fig1}d) show the cospatial photospheric polarity inversion line. Figure~\ref{fig:fig1}f shows the filament as observed using the Interferometric Bidimensional Spectropolarimeter \citep[IBIS;][]{2006SoPh..236..415C} mounted at the Dunn Solar Telescope (DST). In addition to IBIS, observations were taken using the Facility Infrared Spectropolarimeter \citep[FIRS;][]{2010MmSAI..81..763J}, the Rapid Oscillations in the Solar Atmosphere \citep[ROSA;][]{2010SoPh..261..363J}, and the SpectroPolarimeter for Infrared and Optical Regions \citep[SPINOR;][]{2006SoPh..235...55S}. The coud\'e table was rotated such that the slits of FIRS and SPINOR were aligned with the axis of the filament. We focus only on the observations taken with IBIS and FIRS here. Equipped with a high-order adaptive optics system \citep[see][for a detailed description]{2004ApJ...604..906R}, the DST is capable of capturing diffraction-limited images of the Sun at high temporal cadence. The filament was observed at the DST for several days prior to its eruption, occurring on the day following the snapshot shown in Figure~\ref{fig:fig1}. The length of the filament including the additional portion of the filament that extends towards the northern hemisphere (note the filament channel can be seen in the extreme ultraviolet observations) is 660$''$. Due to the total length of the filament of 660$''$ and the limited field of view (FOV) at the DST, only one of the ends of the filament and its surroundings were captured in the ground-based observations.

IBIS scanned across the two lines of H$\alpha$ at 6562.8~\AA\ and Ca {\sc ii} IR at 8542~\AA\ in spectroscopic mode with a non-equidistant spectral sampling of 27 and 30 wavelength points, respectively. The exposure time was 80~ms per image with a total cadence of 13~s for both lines. A circular FOV with a diameter of 95$''$ was sampled with 0.1$''$ pix$^{-1}$ in both x and y. Data were acquired from 13:59:46 to 22:18:10~UT on May 29, with a gap between 15:42:12 and 20:59:00~UT, and from 13:47:08 to 15:02:13~UT on May 30. The IBIS data were reduced with the standard data reduction pipeline\footnote{\href{https://www.nso.edu/wp-content/uploads/2018/05/ibis_tn_005.pdf}{https://www.nso.edu/wp-content/uploads/2018/05/ibis\_tn\_005.pdf}}. In addition to the standard processing, the influence of the pre-filter transmission curve on the line profiles was removed by forcing the average quiet-Sun profile, defined over the region bound by the rectangular box shown in Figure~\ref{fig:fig1}f, into the shape of a reference profile given by the tables of \citet{1961ZA.....53...37D} at $\mu$ = 0.91 \citep[for more details see][]{2019A&A...631A.146S}. The correction retrieved for the average profile was applied to all other spectra. This step also provides an radiometric intensity calibration.

FIRS was used to observe the photospheric Si~{\sc i} 10827~\AA\ line and the chromospheric He~{\sc i} 10830~\AA\ line using a 40-micron wide ($\approx$~0.3$''$), single slit of 75$''$ length that was sampled with 0.15$''$ pix$^{-1}$ along the slit. A spectral range from 10817 to 10856~\AA\ was covered with a spectral sampling of 0.039~\AA\ pix$^{-1}$. The exposure time was 125 ms with a total integration time of 20 s (8 s) for the 4 Stokes parameters per scan step on May 29 (May 30). The noise rms in continuum windows in Stokes I of raw data was 1 $\times$10$^{-2}$ / 3 $\times$10$^{-2}$ / 6 $\times$10$^{-2}$ for the three observations on May 29 and 30. The filament was scanned with 100 (200) steps of 0.3$''$ step width on May 29 (May 30) with a total duration of 40 min (30 min). Data were acquired from 14:41:33 to 15:21:30~UT on May 29 and 13:46:58 to 15:02:14~UT on May 30. The FIRS data were reduced with the standard data reduction pipeline\footnote{\href{https://www.nso.edu/wp-content/uploads/2018/05/firs_soft_manual.pdf}{https://www.nso.edu/wp-content/uploads/2018/05/firs\_soft\_manual.pdf}}. In addition to the standard processing, we ran a two-dimensional Fourier filter to reduce polarized interference fringes and a de-spiking routine to capture hot pixels over the FIRS spectra. Prior to the inversion, the data was re-binned to 0.9$'' \times 0.9''$ per pixel to improve the signal-to-noise ratio. The final noise rms in continuum windows was 8 $\times$10$^{-3}$ of $I_c$ and 4 $\times$10$^{-4}$ of $Q_c$, $U_c$, and $V_c$ for the May 29 and 2 $\times$10$^{-2}$ / 3 $\times$10$^{-2}$ of $I_c$, and 1 $\times$10$^{-3}$ / 1 $\times$10$^{-3}$ of $Q_c$, $U_c$, and $V_c$ for the first / second May 30 FIRS data.

\begin{figure}
\plotone{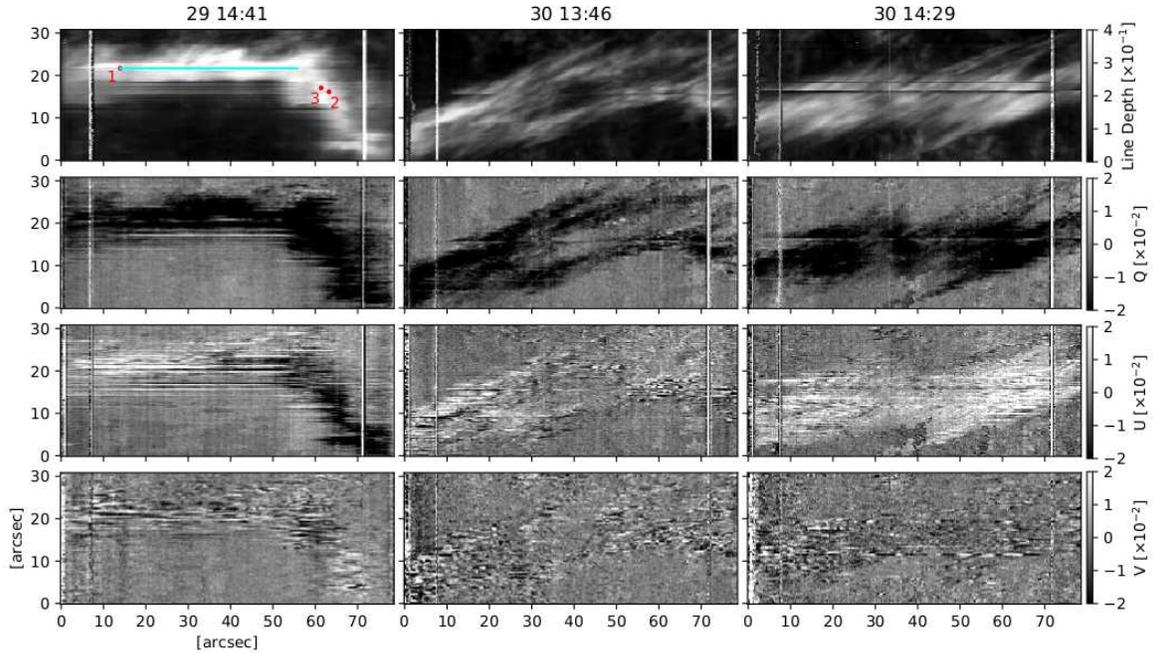}
\caption{Spatial maps from the He~{\sc i} line red core observed by FIRS. From top to bottom: line depth, Stokes Q, U, and V signals integrated over wavelength. Only the red lobe of the two He lines is integrated over wavelength for Stokes V. From left to right: 14:41 on May 29, 2017, 13:46 and 14:29 on May 30. The cyan line in the upper left panel shows the position of the slit spectra in Figure~\ref{fig:fig8}. The three red dots in the upper left panel shows position of Stokes profiles in Figure~\ref{fig:fig3}. \label{fig:fig2}}
\end{figure}

\begin{figure}
\plotone{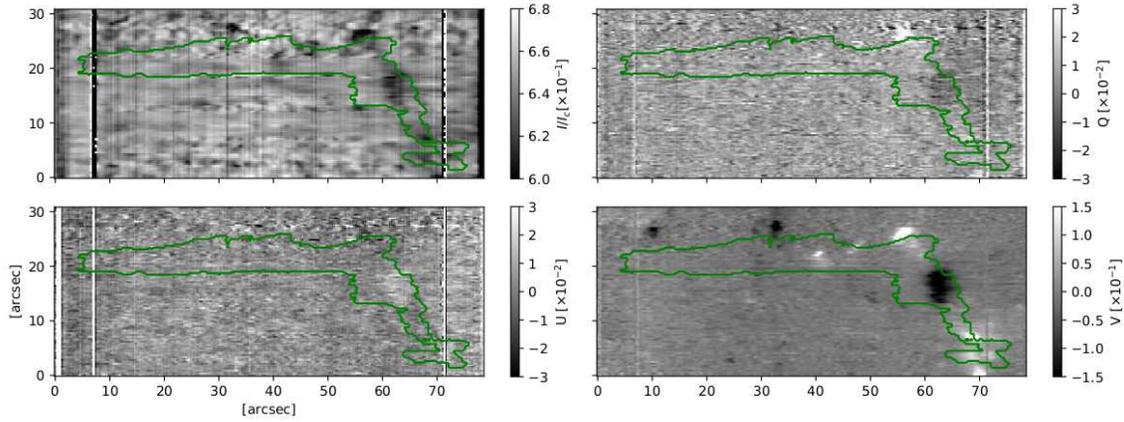}
\caption{Spatial maps from the Si~{\sc i} 10827~\AA\ line observed by FIRS at 14:41 on May 29, 2017. Green contour shows the observed filament in He~{\sc i} 10830~\AA . \label{fig:fig6}}
\end{figure}

Figure~\ref{fig:fig2} displays overview maps of the three FIRS observations in the He {\sc i} line. The filament exhibits a more compact structure on May 29 than on May 30 when several separate threads are visible. The barbs go in the counterclockwise direction, which implies a sinistral filament \citep{1998SoPh..182..107M}. We find significant linear polarization signal inside the filament body in all maps. The circular polarization signal is concentrated towards the end of the filament at the right-hand side of the FOV in these maps on the 29th. The filament was erupting on the 30th. We traced individual threads in the He {\sc i} line depth maps on May 30 as an independent estimate of the direction of the magnetic field azimuth. Figure~\ref{fig:fig6} shows the Stokes images of Si  10827~\AA\ line from the photosphere. The photospheric magnetic field in the field of view is not relevant for the filament apart from the endpoint. The sequence of white-black-white patches at $x\sim 55^{\prime\prime}-70^{\prime\prime}$ corresponds to those found in the HMI magnetogram in Figure \ref{fig:fig1} at $x\sim -380^{\prime\prime} \,-\, -360^{\prime\prime}, y = -70^{\prime\prime}$.

Additional context imagery of the filament and its surroundings was also supplied by the Atmospheric Imaging Assembly \citep[AIA;][]{2012SoPh..275...17L} on the Solar Dynamics Observatory \citep[SDO;][]{2012SoPh..275....3P} and the Extreme Ultra-Violet Imager Telescope \citep[EUVI;][]{2004SPIE.5171..111W} on board the Solar TErrestrial RElations Observatory Ahead \citep[STEREO-A;][]{2008SSRv..136....5K}.

\begin{figure}
\plotone{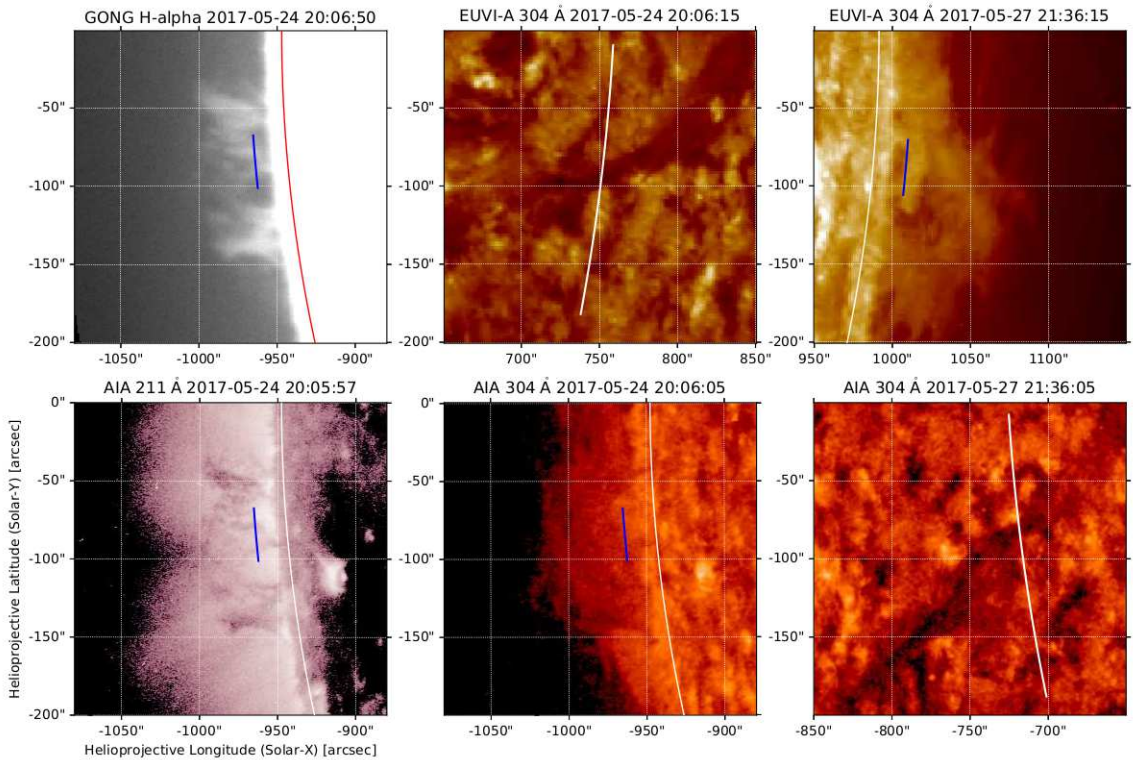}
\caption{The filament observed in emission as prominence at UT 20:06 on May 24, 2017 and UT 21:36 on May 27, 2017. Upper left: GONG H$\alpha$ map at UT 20:06 on May 24, 2017. Lower left: AIA 211~\AA\ map at UT 20:06 on May 24, 2017. Upper middle: STEREO-Ahead/EUVI 304~\AA\ map at UT 20:06 on May 24, 2017. Lower middle: SDO/AIA 304~\AA\ map at UT 20:06 on May 24, 2017. Upper right: STEREO/EUVI 304~\AA\ map at UT 21:36 on May 27, 2017. Lower right: SDO/AIA 304 \AA\ map at UT 21:36 on May 27, 2017. The white line in the upper middle panel shows the limb observed from the earth. The white line in the lower right panel shows the limb observed from the STEREO-Ahead. The red/white lines in the other panels show the position of the solar limb. The blue lines show the height of 20$^{\prime\prime}$ used in the inversion. \label{fig:fig7}}
\end{figure}

\section{Inversion with HAZEL}
\subsection{Inversion Setup}
The HAZEL \citep[HAnle and ZEeman Light;][]{2011ascl.soft09004A} inversion code was used to infer the chromospheric magnetic field vector from the observations. The code requires full Stokes vector data, observation angle, and an initial estimate as input. 

On May 24, the filament appeared at the solar east limb from Earth as shown in the left and middle colomns in Figure~\ref{fig:fig7}. The filament was seen to have spanned between 5$^{\prime\prime}$ and 50$^{\prime\prime}$ above the limb, as observed by both the AIA and GONG instruments, in addition to a clear axis seen at 20$^{\prime\prime}$. At its highest, it extended to a height of about 80$^{\prime\prime}$ above the limb, according to observations taken by AIA in the 304~\AA\ passband, whilst the axis remained at 20$^{\prime\prime}$. On May 27, the filament appeared at the solar west limb when observed by the STEREO-Ahead. This is shown in the upper right panel of Figure~\ref{fig:fig7}; its height remains similar to that observed on May 24. According to these observations the filament was stable between May 24 and May 29, therefore we have used a fixed height of 20$^{\prime\prime}$ above the solar surface in the inversion of the observation on May 29. Based on the line-of-sight velocity information of the two observations on May 30 we have estimated the distance traveled during the eruption process, providing height estimates of 46$^{\prime\prime}$ and 109$^{\prime\prime}$, respectively, for the inversion of the two data sets on May 30. \cite{2008ApJ...683..542A,2013ApJ...768..111S,2019A&A...625A.128D} demonstrated that inversion results are not very sensitive to the specific value of height used, so these estimates are sufficient for our analysis.

\begin{figure}
\plotone{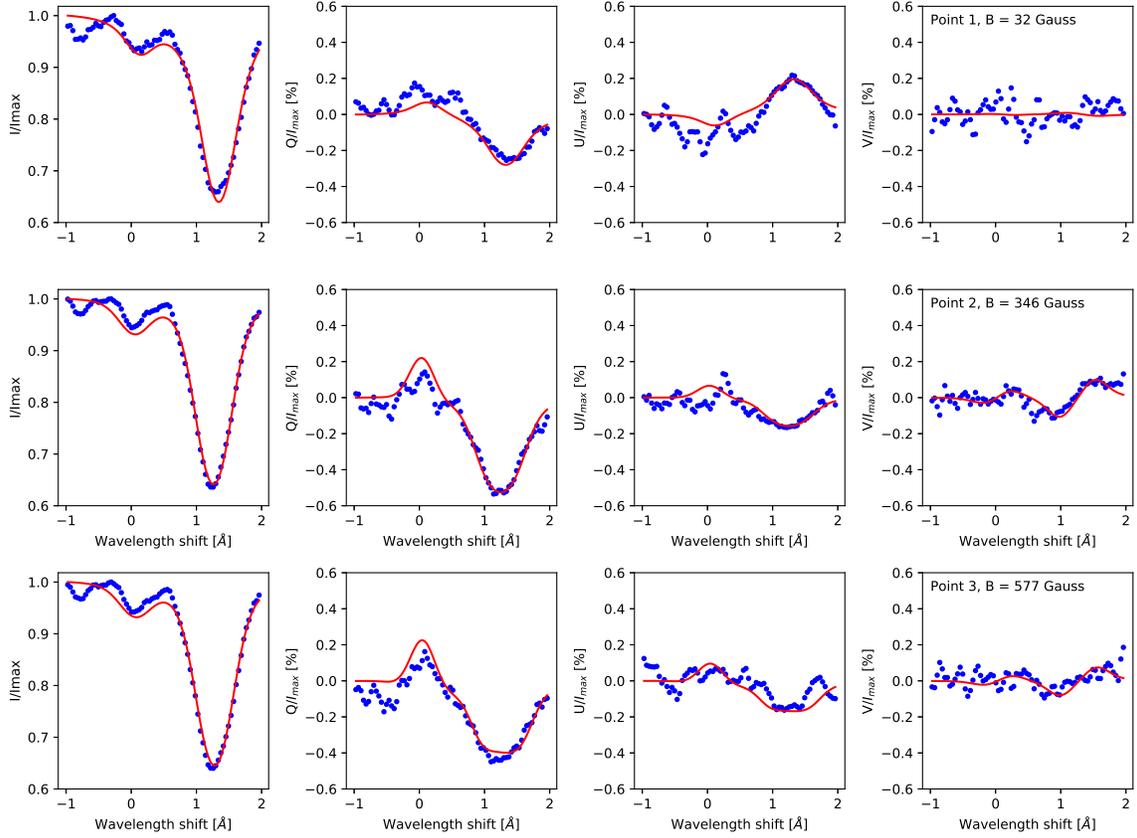}
\caption{Stokes profiles of He~{\sc i} 10830~\AA\ observed on May 29, 2017. Left to right: Stokes I, Q, U and V. The upper/middle/lower panels are correspond to the three numbered positions marked in Figure \ref{fig:fig2} by red dots. Blue dots are the observations. Red curves are best-fit lines from HAZEL. Wavelength shift is from 10829.0911~\AA . \label{fig:fig3}}
\end{figure}

The line profiles from our data in Figure~\ref{fig:fig3} show only one dominant component, examples of the more complex multiple component line profiles can be found in \cite{2011A&A...526A..42S}. As the filament that is the focus of this study is far from any active region and the Stokes V signals are much smaller than the linear polarization signals in most of the filament as shown in Figure \ref{fig:fig2}, we may choose a single slab model for the inversion setup. 

The inversion mode of DIRECT + Levenberg-Marquardt was selected in HAZEL to avoid getting trapped in a local minimum of the merit function. As a result, the initial guess does not have a large influence on the results (see Section 4.2 of \citet{2011ascl.soft09004A} for more details). The seven output parameters of HAZEL are the magnetic field strength B, inclination $\gamma$, azimuth $\Phi$, optical thickness, thermal velocity, damping, and Doppler velocity. We focus here on the magnetic field parameters B, $\gamma$, and $\Phi$. A filament mask was obtained using the dark features in the FIRS Stokes I line-core image of the red component of He~{\sc i}.

\subsection{Disambiguation} 
The inversion provides four different solutions that yield identical spectra because of the 180-degree ambiguity and the Van Vleck ambiguity. The field direction has to therefore be disambiguated using additional a priori information. Two of the solutions included an angle between field lines and the horizontal direction larger than 60 degree. Based on the results of previous studies \citep[e.g.][]{2002Natur.415..403T} in which the field was found largely horizontal with small deviations, we discarded such solutions as unphysical. Furthermore, the leading polarity was positive as shown in the synoptic map of line-of-sight magnetic field (see Figure \ref{fig:fig1}d). Thus, the magnetic field direction along the filament spine is from northwest to southeast \textit{i.e.}, the solution adopted appears to point the axial field in the direction compatible with the observed shear and the tilt angle from Joy's law. \\

\begin{table}
\caption{Root Mean Square of Error and Standard Deviation of Error of Inverted Magnetic Fields} 
\centering 
\small
\begin{tabular}{c c c c} 
\hline\hline 
Observation Time  &  Field Strength [Gauss] & Inclination [degree] & Azimuth [degree] \\ [0.5ex] 
\hline 

2017 May 29 14:41 & $0.5 \pm 0.4$ & $4.7 \pm 2.7$ & $5.8 \pm 3.9$ \\
2017 May 30 13:46 & $2.6 \pm 2.2$ & $11.1 \pm 8.5$ & $15.4 \pm 12.3$ \\
2017 May 30 14:29 & $3.9 \pm 3.5$ & $19.5 \pm 16.5$ & $21.2 \pm 15.0$ \\ [1ex] 
\hline 
\end{tabular}
\label{table:tab1} 
\end{table}

\begin{figure}
\plotone{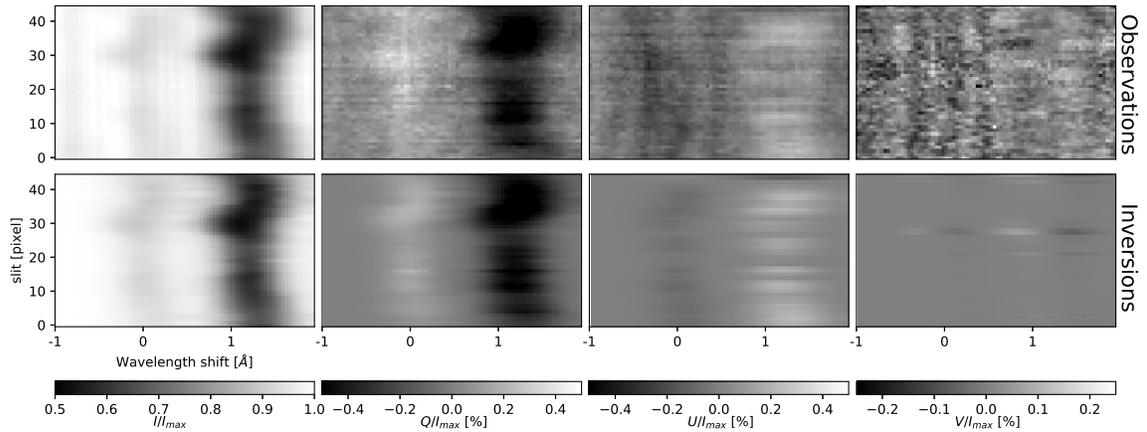}
\caption{Top panels: observed slit spectra of Stokes I,Q,U, and V. Bottom panels: best-fit slit spectra of Stokes I,Q,U, and V. Wavelength shift is from 10829.0911~\AA . \label{fig:fig8}}
\end{figure}

\subsection{Fit Quality}
Figure \ref{fig:fig3} shows the inversion results of one position with strong field and one with weak magnetic field along the filament spine. The upper panels of Figure \ref{fig:fig3} show the inversion results of the pixel at (x: 14$^{\prime\prime}$, y: 22$^{\prime\prime}$) in the spine of the observed filament (red point number 1 on the upper left panel of Figure \ref{fig:fig2}). The inverted field strength, inclination, and azimuth are 32 Gauss, 87 degree, and -120 degree, respectively. The panels in middle row of Figure \ref{fig:fig3} show the inversion results of the pixel at (x: 63$^{\prime\prime}$, y: 16$^{\prime\prime}$) in the spine of the observed  filament (red point number 2 on the upper left panel of Figure \ref{fig:fig2}). The panels in lower row of Figure \ref{fig:fig3} show the inversion results of the pixel at (x: 61$^{\prime\prime}$, y: 17$^{\prime\prime}$) in the spine of the observed  filament (red point number 3 on the upper left panel of Figure \ref{fig:fig2}). The field strength, inclination, and azimuth are 577 Gauss, 91 degree, and -92 degree, respectively. The root mean square of error and standard deviation of error of inverted magnetic field strength, inclination, and azimuth provided by HAZEL are listed in Table \ref{table:tab1}. They are derived from all the 1068 data points. The reversed shape of the Stokes U profiles between the two positions in Figure \ref{fig:fig3} is mainly due to the large difference of azimuth. The signal of Stokes V profiles in the middle and lower panel of Figure \ref{fig:fig3} are stronger than in the upper panel, which indicates a stronger line-of-sight magnetic field. Figure \ref{fig:fig8} shows the fit quality over multiple spectra, and the variation in Q, U, and V along the cyan line in Figure \ref{fig:fig2}. The observations are well reproduced with a single-component model throughout all of the filament. \\

\begin{figure}
\plotone{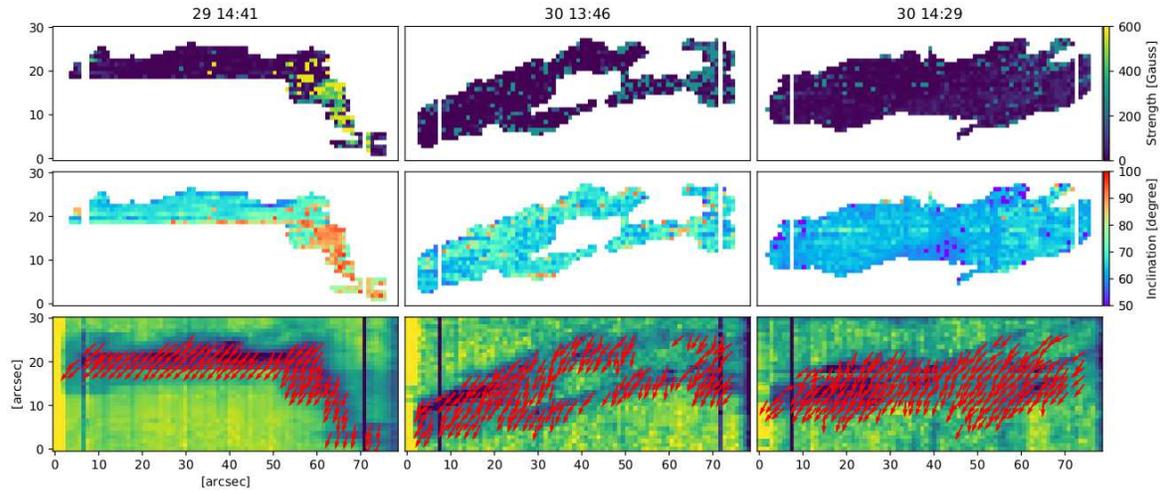}
\caption{Vector magnetic field of the filament observed at 14:41 UT on May 29, 2017. Top row: magnetic field strength. Middle row: inclination. Bottom row: azimuth shown as arrows.The background is a line-core intensity map of He {\sc i}. \label{fig:fig4}}
\end{figure}

\begin{figure}
\plotone{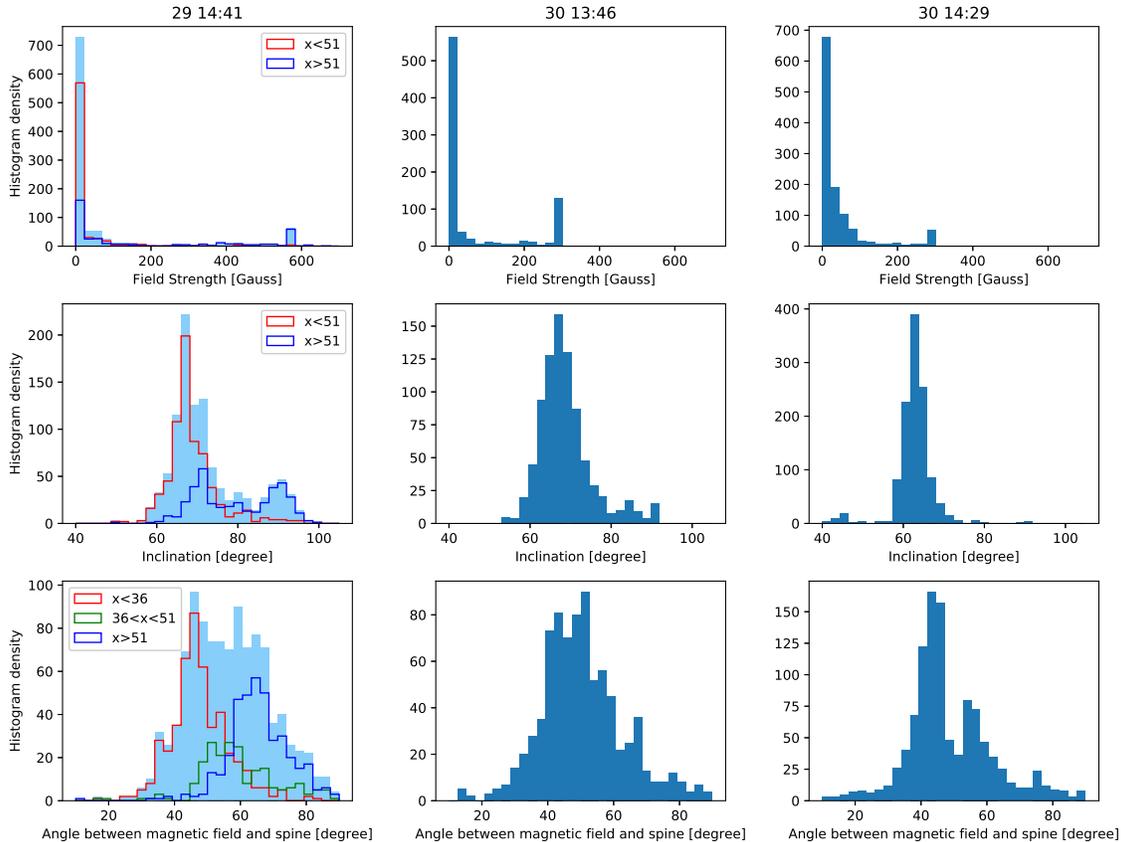}
\caption{Top row: histogram of magnetic field strength. Middle row: histogram of inclination angle of magnetic field. Bottom row: Histogram of angle between magnetic field and spine. From left to right: 14:41 on May 29, 2017, 13:46 and 14:29 on May 30. Positions indicated in the legend of the two left panels refer to positions of Figure \ref{fig:fig4}. \label{fig:fig5}}
\end{figure}

\section{Results} \label{sec:res}

\subsection{Magnetic Field Strength}
Figure \ref{fig:fig4} shows the vector magnetic field of the observed filament. On May 29, the filament was stable. Strong magnetic fields were concentrated at the west end of the filament (x $>$ 51$^{\prime\prime}$) at the right-hand side of the FOV, where the 90th percentile of field strength was 435 Gauss. The mean value at x $<$ 51 arcsec is 24 Gauss. A histogram of the field strength is shown in Figure \ref{fig:fig5}.

The linear polarization of the filament in the He~{\sc i} 10830~\AA\ was strong, while there was almost no linear polarization in the Si~{\sc i} 10827~\AA\ line below the filament body except the end. The structures of the horizontal magnetic fields are visible in the He~{\sc i} layer, but there is no indication in the Si~{\sc i} layer. Unlike linear polarization, the circular polarization at the end of the filament in the He~{\sc i} 10830~\AA\ shows a similar structure as the circular polarization in the Si~{\sc i} 10827~\AA\ below it, and does not show any similarity with the filament shape. The magnetic field strength in the end region of the filament reaches up to 570 Gauss in the HAZEL inversion, while below it in the photosphere, the field strength from HMI is only up to 340 Gauss. Possible reasons for the higher field strength derived from He~{\sc i} are that the photospheric value from the Milne--Eddington inversion of the HMI data is the average over the resolution element because of its spatial and spectral resolution, and filling factor impact. The magnetic field strength at the end is much stronger than in the filament spine, and it is consistant with the assumption that the filament goes down to the photosphere at the end. 

On May 30, the filament started rising at $\sim$ 08:00. The first FIRS observation on this day was taken from 13:46 until 14:20, while the filament was rising. The part of the filament observed by FIRS at this time is shown in Figure \ref{fig:fig1}b as a dark blue contour. The filament width on May 30 was about 12 arcsec, which was larger than its width of 6 arcsec on May 29 (see Figure \ref{fig:fig4}). The filament is not visible in the IBIS H$\alpha$ line-core images on this day, but showed up as a strongly blue-shifted line satellite. On May 30, the mean field strength value for the observation at 13:46 was 70 Gauss. The magnetic field strength was set to be no more than 300 Gauss in the inversion for the observations on May 30. The values of field strength that are close to this limit can be inaccurate due to the decreased line depth of the rising filament and the weak signal of Stokes V for filaments observed close to disk center. The mean optical depth decreased from 0.74 on May 29 to 0.33 on May 30. The rising filament on May 30 was wider than on May 29, and this may be the reason of decreased line depth and optical depth.

Shortly afterwards, a second observation was made starting from 14:29 until 15:02, while the filament was still rising. The mean value of magnetic field strength was 45 Gauss for this observation. The filament disappeared in SDO/AIA 304~\AA\ image at $\sim$ 18:00, and erupted as a minor CME.

\subsection{Angles of Magnetic Field}
The middle row of Figure \ref{fig:fig4} shows the inclination results of the three observations. An inclination value of 0/90/180 is defined to be upward/horizontal/downward with respect to solar surface. On May 29, the inclination angle at positions along the filament spine (x $<$ 51$^{\prime\prime}$) were around 67 degree and similar to each other, while the inclination at the end of the filament was around 79 degree, which is closer to horizontal than the spine region. The histogram of the inclination angles on May 29 is shown in the left panel of Figure \ref{fig:fig5}. The mean values of the magnetic field inclination angle are 80 degree for x $>$ 51 arcsec, and 69 degree for x $<$ 51 arcsec.

The values of inclination are uniform for the observation at 13:46 on May 30 as shown in Figure \ref{fig:fig4}. The mean value of the inclination angle observed at 13:46 is 69 degree. The values of inclination are similar everywhere for the observation at 14:29 on May 30 as shown in Figure \ref{fig:fig4}. The mean value of the inclination angle is 63 degree for this observation.

The Helium threads shown in the upper left panel of Figure \ref{fig:fig2} are aligned with the magnetic field vectors, and they are also aligned with the IBIS H$\alpha$ threads in Figure \ref{fig:fig1}(f).

As shown in Figure \ref{fig:fig4}, the direction of the magnetic field lines of the filament is from the following polarity to the leading polarity. In Figure \ref{fig:fig1}(d), the leading polarity is positive, and the azimuth of photospheric magnetic field is expected to be from the leading polarity to the following polarity. Hence, the inversion results are suggestive of an inverse configuration of the magnetic field direction in the layer observed in He~{\sc i} 10830~\AA .

The lower left panel of Figure \ref{fig:fig5} shows the histogram of the angle between the magnetic field azimuth and the spine axis for the observation of May 29. The mean values of the angle between the magnetic field azimuth and the spine axis are 48 degree for x $<$ 36$^{\prime\prime}$, 59 degree for 36$^{\prime\prime}$ $<$ x $<$ 51$^{\prime\prime}$, and 66 degree for x $>$ 51$^{\prime\prime}$. The angles change gradually along the filament spine for the observation on May 29. This value is 52 degree for the observation at 13:46 on May 30, and was 54 degree for the observation at 14:29 on May 30.

\section{Discussion} \label{sec:dis}

\begin{figure}
\plotone{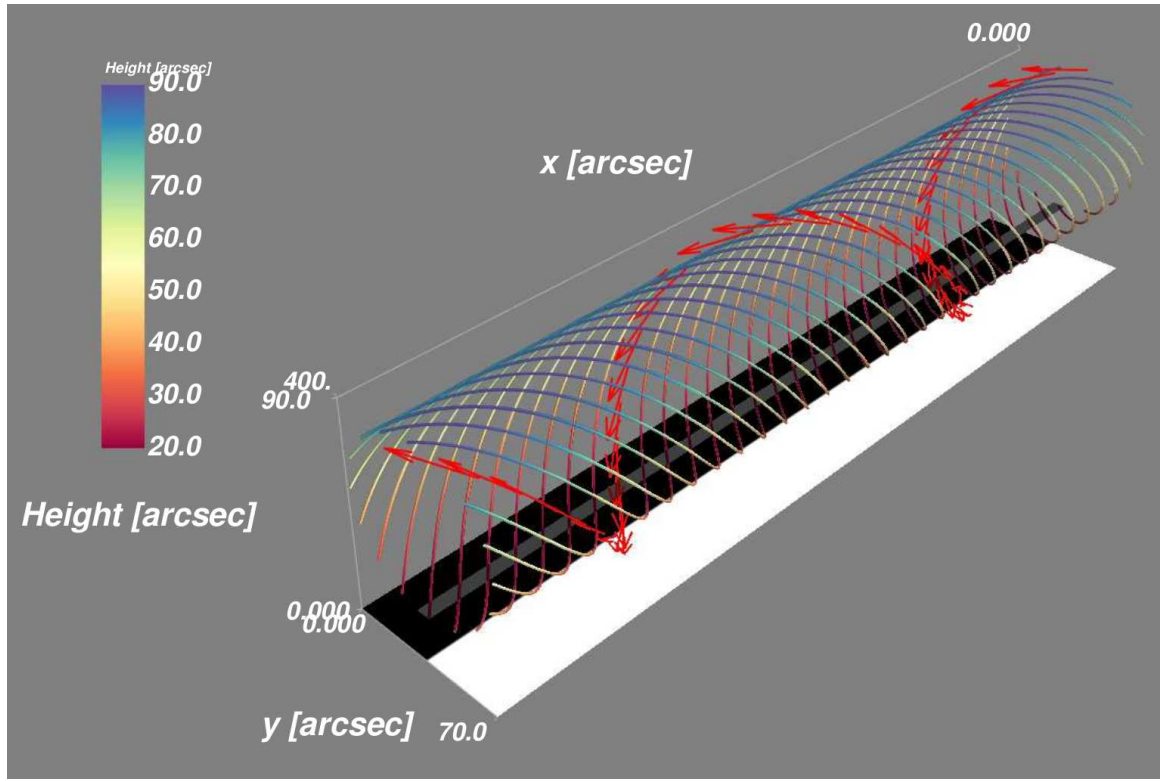}
\caption{Sketch of a filament with flux rope. Colored curves represents magnetic field lines that go through filament. Colors in the magnetic field lines indicate its height, except one line is replaced with a series of red arrows showing the direction of the field. Black/White surface represents polarity of magnetic field in the photosphere under the fluxrope at height 0. Gray surface represents observed surface of filament at the bottom of the fluxrope (height of 20 arcsec). \label{fig:fig9}}
\end{figure}

The magnetic field results are in an inverse polarity configuration indicative of a flux rope topology. By inverse polarity configuration, we mean that the magnetic field vector crossing the polarity inversion line points from negative polarity to positive polarity. The polarity inversion line and photospheric polarities are shown in Figure \ref{fig:fig1} (d). The pointing of magnetic field of the filament is shown as azimuth in the bottom row of Figure \ref{fig:fig4}. Figure \ref{fig:fig9} shows a sketch of this topology based on the observation of both the filament and the photospheric layers on May 29. The flux tube is completely above the photosphere with its bottom at the height of 20 arcsec, where the filament lies. The diameter of the fluxrope in Figure \ref{fig:fig9} is assumed to be 70 arcsec. Thus, each magnetic field line in the sketch makes about two turns (see the red arrows in Figure \ref{fig:fig9}). The filament has a sinistral configuration as expected for the southern hemisphere.

We investigated the vector magnetic field of a quiescent filament from spectropolarimetric scans in the He~{\sc i} 10830~\AA . Our results demonstrate an inverse polarity configuration indicative of a flux rope topology before the eruption process. The magnetic topology does not change during the initial eruption process. 

The magnetic field strength of the filament is mainly 24 Gauss. The value is consistent with studies of quiescent filaments observed at the limb. For example, 20 Gauss from  \cite{2003ApJ...598L..67C}, 7 Gauss from \cite{2014A&A...566A..46O}, and 60 Gauss from \cite{2015ApJ...802....3M}. However, the magnetic field strength at the filament end is much stronger than the value at the central part, and the same trend is shown in studies at the limb \citep{2014A&A...566A..46O}. The structures of the horizontal magnetic fields are visible in the He~{\sc i} layer, but there is no indication in the Si~{\sc i} layer. In contrast, AR filaments are usually observed with large scale horizontal structures in both the He~{\sc i} layer and the Si~{\sc i} layer \citep{2012ApJ...749..138X,2012A&A...539A.131K,2019A&A...625A.128D}. The magnetic field in the filament body seems completely detached from the photospheric field below it, and it is in line with the hyperbolic flux tube \citep{2002JGRA..107.1164T,2003ApJ...582.1172T} interpretation of a flux rope, whereas AR filaments are believed to be low-lying and therefore contain bald point separatrix surface signatures in the photosphere.

The inclination angle to the solar vertical is around 67 degree for the observed part of the stable filament on May 29. Reported inclination of quiescent prominence ranges widely from mostly horizontal \citep{2003ApJ...598L..67C} to 30 degree from vertical \citep{2015ApJ...802....3M}, such as 77 degree from vertical \citep{2014A&A...566A..46O}. The angle between magnetic field azimuth and filament spine is 47 degree to 58 degree, and gradually decreases along the filament from the end on May 29. \cite{2014A&A...566A..46O} reported 58 or 24 degree for the angle of a  quiescent prominence. In general, the polarimetric data can be used to determine the chirality of filaments without information of barb direction.

The flux rope topology is found on the day prior to the eruption, and during the eruption without any evident changes. The filament was observed on two consecutive days, and was rising while being observed on the second day. The inversions give the field strength for both days. During the eruption, the observed magnetic structure is well preserved without significant changes. The eruption itself does not seem to impact the observed field strengths significantly.

The quiescent filament has a length of about 660 arcsec, and a width of about 6 arcsec in the H$\alpha$ observation on May 29, 2017. The filament expanded with its width doubled on the second day and the line depth decreased as shown in Figure \ref{fig:fig2}. The line depth became less on May 30, 2017 due to rising of the filament. These facts are consistent with the rise and expansion of the filament through the solar atmosphere. The filament spine was almost straight with both ends in S-shape. It resemble the sigmoidal morphology observed in X--ray of Active Regions \citep{1999GeoRL..26..627C}.

\section{Conclusions} \label{sec:con}
We derived the magnetic field vector in a filament prior to and during its eruption process. The magnetic field topology indicates a flux rope. The field topology does not change significantly during the initial phase of the eruption. Using this information on the magnetic field of erupting material on the Sun could improve the prediction of the geoeffectiveness of the related solar storms. The synoptic observations that are currently run at the DST aim to provide suited observations for this purpose.

This work is funded by NSF 1839306. Sunspot Solar Observatory is a multi institution consortium that is funded by multiple entities including NSF (1649052, 1945705) and the State of New Mexico. Funding for the DKIST Ambassadors program is provided by the National Solar Observatory, a facility of the National Science Foundation, operated under Cooperative Support Agreement number AST-1400405. J.M.J. thanks the STFC for support via funding given in his PhD Studentship, and travel funds awarded by the Royal Astronomical Society. D.M.L. acknowledges support from the European Commission's H2020 Programme under the following Grant Agreements: GREST (no. 653982) and Pre-EST (no. 739500) as well as support from the Leverhulme Trust for an Early-Career Fellowship (ECF-2014-792) and is grateful to the Science Technology and Facilities Council for the award of an Ernest Rutherford Fellowship (ST/R003246/1). D.P.C. was partially supported through NSF grant AGS-1413686. K.M. acknowledges support by the NASA Heliophysics Guest Investigator program and the NASA cooperative agreement NNG11PL10A.

\bibliographystyle{unsrt}

\end{document}